\documentclass[preprint,showpacs,preprintnumbers,amsmath,amssymb]{revtex4}

\usepackage{etex}
\usepackage{graphicx}
\usepackage{dcolumn}
\usepackage{bm}
\usepackage{amssymb,amsthm,amscd,amsbsy,array}
\usepackage{amsmath,dsfont}
\usepackage{soul} 
\usepackage{graphics,xcolor}

\newcommand{\half}{{\frac{1}{2}}}
\def\2{{\half}}
\newcommand{\const}{\mathop{\rm const}\nolimits}
\def\parag{\hfil\break} 
\def\kikezd{\parag\underbar}
\def\bA{{\bm{A}}}

\def\bp{{\bm{p}}}

\def\ba{{\bm{a}}}

\def\bnabla{\mbox{\boldmath$\nabla$}}
\def\bTheta{\mbox{\boldmath$\Theta$}}
\def\br{{\bm{r}}}

\def\bE{{\bm{E}}}
\def\bB{{\bm{B}}}

\def\bnabla{{\bm{\nabla}}}

\def\bp{{\bm{p}}}

\def\hbp{{\widehat{\bm{p}}}}
\def\bx{{\bm{x}}}
\def\bz{{\bm{z}}}
\def\by{{\bm{y}}}

\def\beq{\begin{equation}}
\def\eeq{\end{equation}}
\def\beqa{\begin{eqnarray}}
\def\eeqa{\end{eqnarray}}

\def\barray{\left(\begin{array}}
\def\earray{\end{array}\right)}
\def\barraynb{\begin{array}}
\def\earraynb{\end{array}}
\def\benu{\begin{enumerate}}
\def\eenu{\end{enumerate}}

\def\Ort{{\rm O}}
\def\smallover#1/#2{\hbox{$\textstyle\frac{#1}{#2}$}} %

\newcommand{\fm}{{\mathfrak{{m}}}}

\newcommand{\cE}{{\mathcal{E}}}

\usepackage{color}
\newcommand{\red}{\textcolor{red}}  

\newcommand{\green}{\textcolor{green}}

\begin{document}

\preprint{arXiv:1409.4225v2
}

\title{Anomalous Hall Effect for semiclassical chiral fermions \\
}

\author{
P.-M. Zhang $^{1}$
\footnote{e-mail:zhpm@impcas.ac.cn},
P. A. Horv\'athy $^{1,2}$\footnote{e-mail:horvathy@lmpt.univ-tours.fr}
}

\affiliation{
\\
$^1$Institute of Modern Physics, Chinese Academy of Sciences, Lanzhou, (China)
\\
${}^2$Laboratoire de Math\'ematiques et de Physique
Th\'eorique,
Universit\'e de Tours,  
(France) 
}

\date{\today}

\begin{abstract}
Semiclassical chiral fermions manifest the anomalous spin-Hall effect: when put into a pure electric field they suffer a side jump, analogous to what happens to their massive counterparts in non-commutative mechanics. The transverse shift is consistent with the conservation of the angular momentum. In a pure magnetic field instead,  spiraling motion is found. Motion in Hall-type perpendicular electric and magnetic fields is also studied. 
\end{abstract}


\maketitle



Semiclassical massless chiral fermions have attracted considerable recent interest
\cite{SonYama,Stephanov,hadrons,Dunne,SonYama2,ChenWang,Stone,Chen2013,ChenSon,Manuel,NewStone,DHchiral,Karabali,Jho}.
The model in \cite{Stephanov}, for example, describes a spin-$1/2$ system with positive helicity and energy by the phase-space action
\beq
S=\int\Big(\big(\bp+e\bA\big)\cdot\frac{d{\bx}}{dt}-\big(|\bp|+
e\phi
\big)
-\ba\cdot\frac{d{\bp}}{dt}
\Big)dt,
\label{chiract}
\eeq
which also involves
an the additional ``momentum-dependent vector potential'' $\ba(\bp)$ for the ``Berry monopole'' in $\bp$-space \cite{Niu},
$ 
\bnabla_{\bp}\times\ba=\bTheta\equiv
\frac{\hbp}{2|\bp|^2}\,,
$ 
 where $\hbp$ is the unit vector $\hbp=\displaystyle{\bp}/{|\bp|}$. Here $\bA(\bx)$ and $\phi(\bx)$ are ordinary vector and scalar potentials and $e$ is the electric charge. Variation of the chiral action (\ref{chiract}) yields the equations of motion for position $\bx$ and momentum $\bp\neq0$ in three-space, 
\beq\left\{\barraynb{lll}
\fm 
\,\displaystyle\frac{d{\bx}}{dt}&=&\hbp+e\,{\bE}\times\bTheta+(\bTheta\cdot\hbp)\,e\,\bB,
\\[12pt]
\fm \,\displaystyle\frac{d{\bp}}{dt}&=&e\,\bE+{e}\,{\hbp}\times\bB+e^2(\bE\cdot\bB)\,\bTheta,
\earraynb\right.
\label{chireqmot}
\eeq
 where  $\bE$ and $\bB$ are the electric and magnetic field, respectively, and
$\fm= 1+e\,\bTheta\cdot\bB$
is an effective mass.  
 
These equations are strongly reminiscent of those in non-commutative mechanics \cite{Niu,DHHMS}, which allowed us to propose a simple mechanical model for the  the Anomalous Hall Effect, i.e., the transverse shift  observed in some ferromagnetic materials in the absence of any magnetic field  \cite{Karplus,AHE}.
 
The aim of this Note is to prove a similar
 result  also for massless fermions.
  
 \goodbreak



\begin{figure}[ht]\vskip-2mm
\begin{center}
\includegraphics[scale=.42]{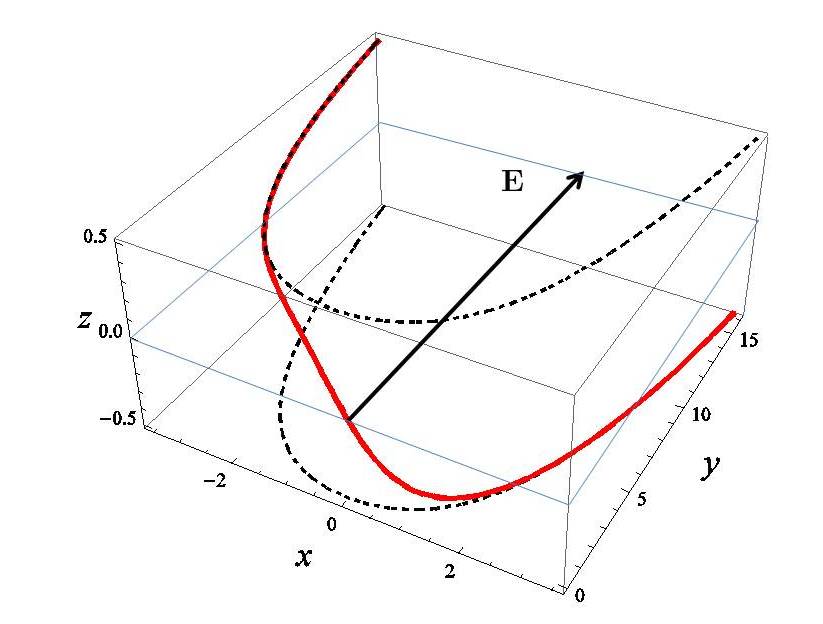}
\vspace{-8mm}
\end{center}
\caption{\it A chiral particle exhibits the \underline{anomalous Hall effect}, when put into a pure electric field. For large values of $|t|$ the motion is essentially a chain curve in the planes ${z}(\pm\infty)$, but as $t\to0$ it suffers a \underline{transverse shift} 
 perpendicular to $\bE$ and the initial momentum, jumping from from the initial plane ${z}(-\infty)$ to 
${z}(\infty)$.
} 
\label{figAHE}
\end{figure}

(i) Let us first study what happens within the chiral model
in a \emph{constant pure electric field},
$\bB=0,\, \bE=E\hat{\by},\, E=\const$. The equations of motion (\ref{chireqmot}) then become,
\beq
\frac{d\bx}{dt}=\hbp+e\bE\times\bTheta,
\qquad
\frac{d\bp}{dt}=e\bE\,.
\label{StephEeq}
\eeq
Thus $\bp(t)=\bp_0+e\bE\,t$. Choosing, for example, the initial momentum ${\bp}_0=eE\hat{\bx}$ and  $\bx=0$
for initial position, we get,
\beq
x(t)={\rm arg sh}\, t,\qquad
y(t)=\sqrt{1+t^2}-1,\qquad
z(t)=-\frac{1}{2eE}\frac{t}{\sqrt{1+t^2}}
\eeq
cf. Fig. \ref{figAHE}. 
The remarkable result is that, \emph{due to the anomalous velocity term in eqn. (\ref{StephEeq}), the particle follows a $3D$ trajectory even for planar initial conditions}: for large values of $|t|$,  the motion is approximately a chain curve $y(x)=\cosh x-1$ (as it would be for $\bTheta=0$) lying, for $t=\mp\infty$, in the planes ${z}(\pm\infty)=\mp(2eE)^{-1}$, respectively.
With $|t|$ approaching $0$, however, the trajectory abruptly leaves the initial plane of the motion and suffers a \underline{transverse shift} perpendicular to $\bE$ and the initial momentum,
\beq
\Delta z = z(\infty)-z(-\infty)=-\frac{1}{eE}\,.
\label{Eshift}
\eeq
Thus, the chiral model provides a (semi)classical description of the \emph{anomalous Hall effect}.

A similar behavior was observed before for
the non-relativistic dispersion relation $\cE=\bp^2/2$ \cite{AHE}.
It is also reminiscent of the ``side jump''  for the scattering of free chiral fermions
\cite{ChenSon}, as well as of the optical Hall effect \cite{OHE}.

The shift formula (\ref{Eshift}) is consistent with the conservation of the angular momentum.
The constant electric field  breaks the full rotational symmetry to  $\Ort(2)$ of rotations around the direction (chosen to be $\hat{\by}$) of $\bE$; [$E$ times] the associated angular momentum is 
\beq
\ell\equiv
\bE\cdot(\bx\times\bp)+\frac{\bE\cdot\bp}{2|p|}.
\label{Eangmom}
\eeq
Then $\ell=eE^2z(t)+\displaystyle\frac{Et}{2\sqrt{1+t^2}}=0$ for $t=0$
and thus for any $t$; therefore $eE\Delta z= -1$.


\vskip2mm
(ii) Let us assume  instead that we have a \emph{constant pure magnetic field},
$\bE=0$ and $\bB=B\hat{\bz}$, $B=\const$. Then the chiral eqns. of motion (\ref{chireqmot}) reduce to 
\beq\left\{\barraynb{lll}
\fm \,\displaystyle\frac{d{\bx}}{dt}&=&\hbp+\displaystyle\half\frac{e\bB}{|\bp|^2}
\\[12pt]
\fm 
\,\displaystyle\frac{d{\bp}}{dt}&=&{e}\,{\hbp}\times\bB
\earraynb\right.
\qquad\hbox{where}\qquad
\fm=1+e\,\displaystyle\frac{\hbp\cdot\bB}{2|\bp|^2}\,.
\label{pureBchireqmot}
\eeq
 $|\bp|$,  $\hbp\cdot\bB=
B\cos\vartheta$ and $p_z=\bp\cdot\hat{\bz}$ and thus also $\fm$ 
are constants of the motion.

If the effective mass does not vanish, $\fm \neq0$, then $\bp$ precesses around the ${\bz}$-axis with angular velocity $\omega=-eB/|\bp|\fm$,
$ 
\bp(t)=(p_0e^{-i(eB/|\bp|\,\fm)t},p_z)
$ 
where $p_0={p_0}_x+i{p_0}_y$.
The anomalous term in the upper relation merely  adds to the drift along the $\bB$ direction,
\beq
\bx(t)=\Big(\frac{ip_0}{eB}e^{-ie(B/|\bp|\fm)t},z(t)\Big),
\qquad
z(t)=\big(\frac{\cos\vartheta}{\fm}+\frac{eB}{2|\bp|^2\fm}\big)t+z_0.
\label{Stephx}
\eeq 
\begin{figure}[h]
\begin{center}\vskip-3mm
\includegraphics[scale=.4]{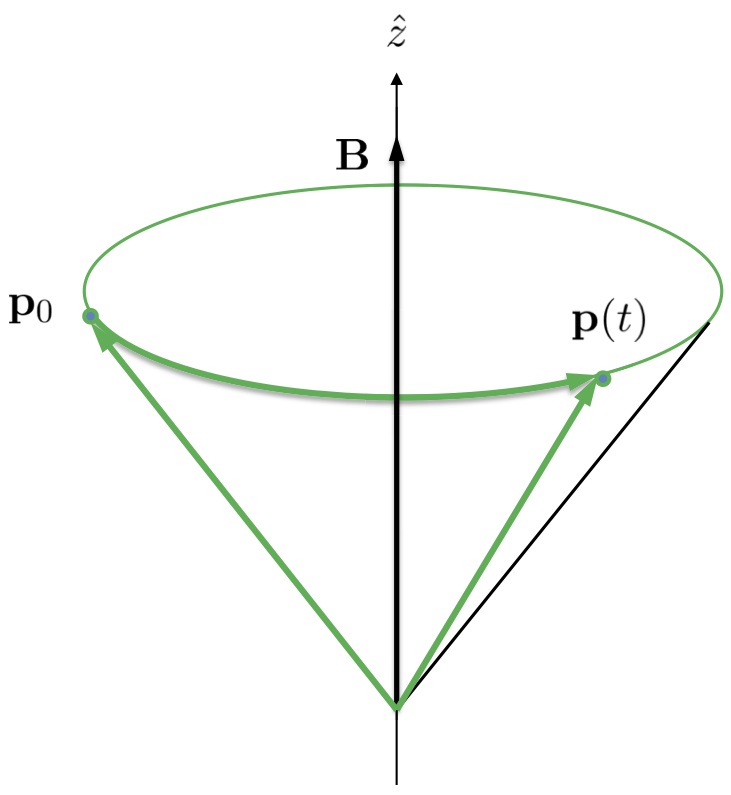}
\qquad
\includegraphics[scale=.46]{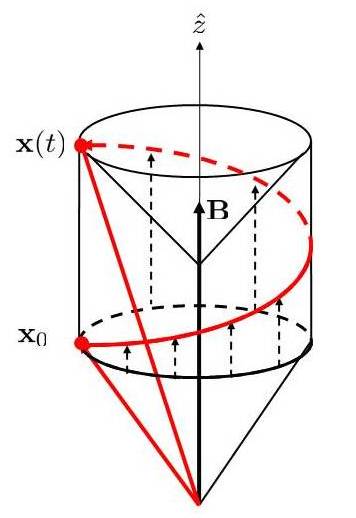}\\
\hskip10mm (a) \hskip49mm (b)
\end{center}\vspace{-6mm}
\caption{\it (a) Motion in pure constant  magnetic field $\bB$. The momentum $\green{\bp}(t)$  precesses around the $\bB$-direction. (b) The position $\red{\br}(t)$ spirals on a cylinder around the $\bB$-axis, obtained by combining precession with the vertical drift of the supporting cone itself.}
\label{figAHE2}
\end{figure}
This cork-screw-like spiraling motion is reminiscent of the one found for heavy ions \cite{Dunne}.

For the sake of comparison, we record [$B$-times] the conserved angular momentum along the magnetic field,
\beq
\ell=\bB\cdot\bx\times\bp+\half\bB\cdot\hat{\bp}+\frac{e}{2}(\bB\times\bx)^2.
\label{Bangmom}
\eeq
The presence of the last term here  is required by the equations of motion (\ref{pureBchireqmot}).

When $|\bp|^2=-(eB/2)\cos\theta$, then the effective mass vanishes, $\fm=0$, and the system becomes singular;  it requires reduction, as does its planar counterpart \cite{DHexo}.
Eqn (\ref{pureBchireqmot}) implies that the momentum is vertical, $\theta=0$ [and therefore $p_z^2=-(eB/2)$, which in turn requires $eB<0$]. Then the upper eqn. is identically satisfied, leaving $\bx(t)$ undetermined. 

(iii) Let us consider what happens when the fields are combined. For simplicity, we
only consider \emph{perpendicular electric and magnetic fields},
$
\bE=E\hat{\by}, \,\bB=B\hat{\bz},
$ 
when a massive particle would perform Hall motion.
The eqns of motion read
\beq
\left\{\barraynb{lll}
\fm\dot{x}&=&\displaystyle\frac{p_{x}}{|\bp|}+eE\displaystyle\frac{p_{z}}{2|\bp|^{3}}  
\\[8pt]
\fm\dot{y}&=&\displaystyle\frac{p_{y}}{|\bp|}
\\[8pt]
\fm\dot{z}&=&\displaystyle\frac{p_{z}}{|\bp|}-eE\displaystyle\frac{p_{x}}{2|\bp|^{3}}+\displaystyle\frac{eB}{2|\bp|^{2}}  
\earraynb\right.
\qquad
\left\{\barraynb{lll}
\fm\dot{p}_{x}&=&\;\; eB\displaystyle\frac{p_{y}}{|\bp|}
\\[8pt]
\fm\dot{p}_{y}&=&-eB\displaystyle\frac{p_{x}}{|\bp|}+eE
\\[8pt]
m\dot{p}_{z}&=&0 
\earraynb\right.
\quad
\fm=1+\displaystyle\frac{eBp_{z}}{2|\bp|^{3}}\,.
\label{EBeqmot}
\eeq 
Therefore the $p$-diagram (analogous to the hodograph) is 
a curve in the horizontal plane $p_z=\const$.
Combining the other eqns, 
\begin{equation}
\frac{dp_{x}}{dt}=eB\frac{dy}{dt}
\quad\Rightarrow\quad
p_{x}=p_{x0}+eB\left(y-y_{0}\right) .  
\label{pysol}
\end{equation}%
Then, imitating the elementary derivation of energy conservation, we multiply the $p$-equations with
$\dot{\bp}$ to infer
\begin{equation}
\frac{1}{2}\fm\frac{d}{dt}\left(|\bp|^{2}\right) =eEp_{y}.
\end{equation}%
Re-inserting into (\ref{EBeqmot})
 yields
\begin{equation}
\frac{d|\bp|}{dt}=\frac{E}{B}\,\frac{dp_{x}}{dt}
\quad\Rightarrow\quad
|\bp|=\frac{%
E}{B}\left(p_{x}-p_{x0}\right)+|\bp|_{0}.  
\label{psol}
\end{equation} 
Dividing $\dot{p}_x$ by $\dot{p}_y$
allows us to deduce the ``$p$-hodograph'',
\begin{equation}
\left(B^{2}-E^{2}\right) p_{x}^{2}
-2E\left(B|\bp|_{0}-Ep_{x0}\right) p_{x}
+B^{2}p_{y}^{2}=\left( Bp_{0}-Ep_{x0}\right) ^{2}-B^{2}p_{z0}^{2}, 
\label{pxysol2}
\end{equation}%
where $|\bp|_{0},~p_{x0}$ are constants of integration.
Eqn. (\ref{pxysol2}) describes a \emph{conic section}, namely an ellipse/parabola/hyperbola, depending on $|E|$ being smaller/equal/larger as $|B|$. Bounded $p$-hodographs  arise therefore in strong magnetic fields, $|E|<|B|$.

Coming to motion in real space, from (\ref{pysol}) we infer that
\begin{equation}
y(t)=\frac{1}{eB}\left(p_{x}(t)-p_{x0}\right)+y_{0},
\end{equation}%
so that $y(t)$oscillates when $p_{x}(t)$ does so,
i.e., for $|B|>|E|$. 

For the sake of further simplification, we assume henceforth that $p_z=0$; then
\begin{equation}
x(t)=\frac{E}{B}t-\frac{1}{eB}(p_{y}(t)-p_{y0})+x_{0},
\end{equation}%
which is a \emph{Hall-type motion perpenducularly to both the electric and magnetic fields}, combined with that of $p_y(t)$. This motion is thus unbounded, except in the purely magnetic case $E=0$.

\begin{figure}[h]
\begin{center}
\includegraphics[scale=.4]{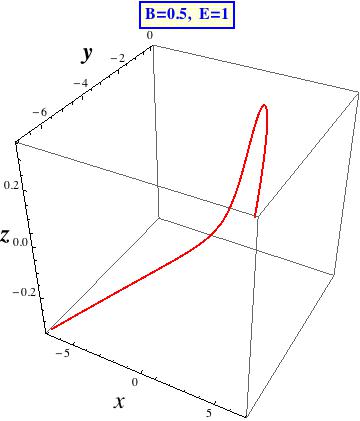}\,
\includegraphics[scale=.39]{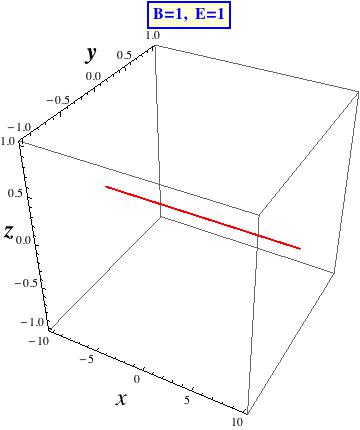}\,
\includegraphics[scale=.38]{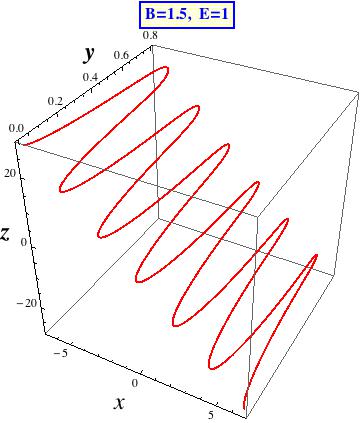}
\vspace{-8mm}
\end{center}
\caption{\it Motion in  perpendicular electric and magnetic fields combined as in Hall effect. On all our plots we took $\bE=E\hat{\by},\,\bB=B\hat{\bz}$ and initial conditions  $\bp(0)=\hat{\bx}$ and $\bx=0$.
The plots of Fig. \ref{EBsolfig+} correspond to $B>0$.}
\label{EBsolfig+}
\end{figure}

\begin{figure}[h]
\begin{center}
\includegraphics[scale=.385]{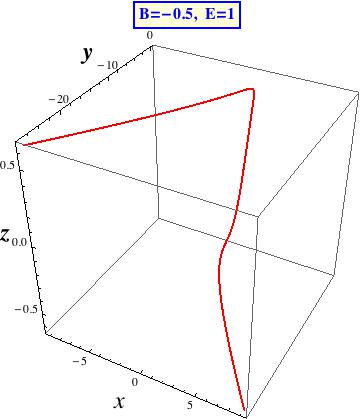}\,
\includegraphics[scale=.38]{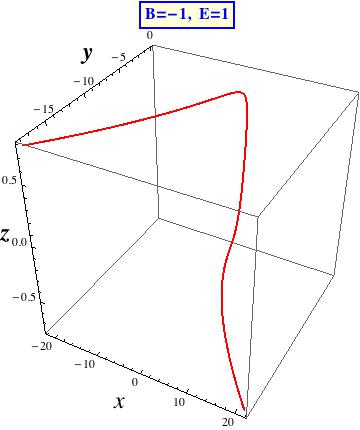}\,
\includegraphics[scale=.38]{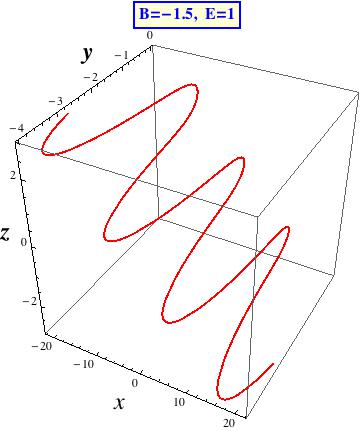}
\vspace{-8mm}
\end{center}
\caption{\it Motion in  perpendicular electric and magnetic fields
as in Fig. \ref{EBsolfig+}, but with $B<0$.}
\label{EBsolfig-}
\end{figure}

For the $z$-eqn,
\begin{equation}
\dot{z}=-eE\frac{p_{x}}{2|\bp(t)|^{3}}
+\frac{eB}{2|\bp(t)|^{2}}\,,
\label{zdot}
\end{equation}
we only found numerical solutions, 
shown on Figs \ref{EBsolfig+} and \ref{EBsolfig-}. 

\goodbreak
Finite transverse shift, $\Delta z<\infty$, requires the
electric field to dominate, $|E|\geq |B|$, since then $|\bp|$ is unbounded by (\ref{pxysol2}), so that  $\dot{z}\to0$ by (\ref{zdot}). But the value of the initial momentum plays a role also, and we have not been able to derive an exact formula like (\ref{Eshift}), let alone a precise threshold between bounded or unbounded transverse shifts. 

Both numerical and analytical study confirm that for $B\to0$ and for $E=0$, respectively,  the purely electric and magnetic cases are recovered as expected.

\kikezd{Conclusion}

 In this Note, we studied the motion of a  semiclassical massless chiral fermion in a constant electromagnetic field. In a dominating electric field the system suffers a finite transverse shift $\Delta z < \infty$ as in the \emph{anomalous Hall effect} \cite{Karplus}, and similar to what was found before in non-commutative mechanics \cite{AHE}. If the magnetic field dominates, the transverse shift becomes unbounded, $\Delta z = \infty$. The  motion parallel to $\bE$  is oscillatory when $\bB^2-\bE^2>0$ and unbounded when $\bB^2-\bE^2\leq0$. Perpendicularly to both $\bE$ and $\bB$ the motion  is unbounded except in a pure magnetic field, when it spirals as shown in Fig. \ref{figAHE2}.

We mention that the same problem can also be studied within the framework proposed in \cite{DHchiral}.
The motions are \emph{substantially different}, underlining the importance of the coupling rule one adopts \cite{DHZ}.


\begin{acknowledgments} 
We would like to thank Mike Stone for calling our attention at chiral fermions and to Christian Duval 
 for stimulating discussions and correspondence. 
\end{acknowledgments}

\goodbreak


\end{document}